\documentclass{article}
\usepackage{graphicx,psfig,epsfig}
\def\title{\begin{center}\Large\bf}
\def\author(s){\vspace{0.3cm}\large\rm}
\DeclareRobustCommand{\ion}[2]{%
\relax\ifmmode
\ifx\testbx\f@series
{\mathbf{#1\,\mathsc{#2}}}\else
{\mathrm{#1\,\mathsc{#2}}}\fi
\else\textup{#1\,{\mdseries\textsc{#2}}}%
\fi}
\newcommand{\ha}{\rm H$\alpha$}
\newcommand{\hbeta}{\rm H$\beta$}

\newcommand{\sii}{$[$\ion{S}{ii}$]$}
\newcommand{\oii}{$[$\ion{O}{ii}$]$}
\newcommand{\oiii}{$[$\ion{O}{iii}$]$}

\def\degr{\hbox{$^\circ$}}
\newcommand{\snr}{G 65.3+5.7}
\newcommand{\vel}{km s$^{-1}$}
\def\text{\end{center}}

\begin{document}

\title
A CCD mosaic of an extended remnant in Cygnus

\author(s)
 F. Mavromatakis$^{\rm 1}$, P. Boumis$^{\rm 1}$, and E.V. Paleologou$^{\rm 2}$

$^{\rm 1}${\it University of Crete, P.O Box 2208, GR-71003, Heraklion, Greece}\\
$^{\rm 2}${\it Foundation for Research and Technology-Hellas, P.O. Box 1527,\\
GR-71110 Heraklion, Greece}\\
\text

\vspace{0.3cm}

\large

\section*{Abstract}
We present the first CCD mosaic ever produced for the supernova remnant 
G 65.3+5.7 in \oii\ 3727 \AA\ at a moderate angular resolution.  
The remnant was observed with the 0.3 m wide--field telescope at 
Skinakas Observatory, Crete, Greece. 
Low resolution spectroscopy was performed at selected areas around 
this extended remnant. 
The spectral data are under analysis, while a first processing of 
the imaging observations produced impressive results.

\section{Introduction}
Gull et al. (1977) reported the detection of a new 
supernova remnant during an emission line survey of the Milky Way. 
The remnant is found in Cygnus and is located a few degrees to the 
south--west of CTB~80. 
The wide field covered by this remnant was imaged on plates by the 
authors in the emission lines of \oii\ and \ha. 
Fesen et al. (1983, 1985) performed higher resolution imaging 
observations in \oiii 5007 \AA\ of five (5) selected portions of the remnant and 
spectral observations at two (2) positions. 
Major characteristics of the remnant are its large extent ($\sim$5\degr $\times$
4\degr), and its filamentary appearance in the medium ionization line of 
\oiii.
The remnant is also detected in the radio survey of Reich et al. (1979), 
while the Einstein IPC mosaic shows very weak X--ray emission (Seward 1990). 
An average distance of $\sim$ 1 kpc to this remnant is given in the 
literature.  

\section{Observations}

The imaging observations were performed with a 1024 $\times$ 1024 Site CCD 
which in conjuction with the 0.3 m telescope at Skinakas Observatory 
resulted in a pixel scale of 5 arcseconds. Each field was observed for 
1800 sec with this highly efficient CCD camera in \oii\ and \oiii. 
The astrometric solutions were calculated with the aid of the Hubble Space 
Telescope Guide Star catalogue for all individual fields and then they were 
projected to a common origin for further processing.  
\par
The long--slit spectral observations were conducted with a 1300 line mm$^{-1}$ 
grating and a 800 $\times$ 2000 Site CCD camera. This camera was mounted on the 
1.3 m telescope at Skinakas Observatory and the range of 4750 \AA\ -- 6815 \AA\ 
was covered at a scale of $\sim$ 1 \AA\ per pixel.
The total exposure times are 7800 sec and standard IRAF and MIDAS routines 
were employed for the reduction of the data. The spectra are flux 
calibrated through the observations of several spectrophotometric standard
stars.
\begin{figure}
\centering
\mbox{\epsfclipon\epsfxsize=2.8in\epsfbox[72 161 540 631]{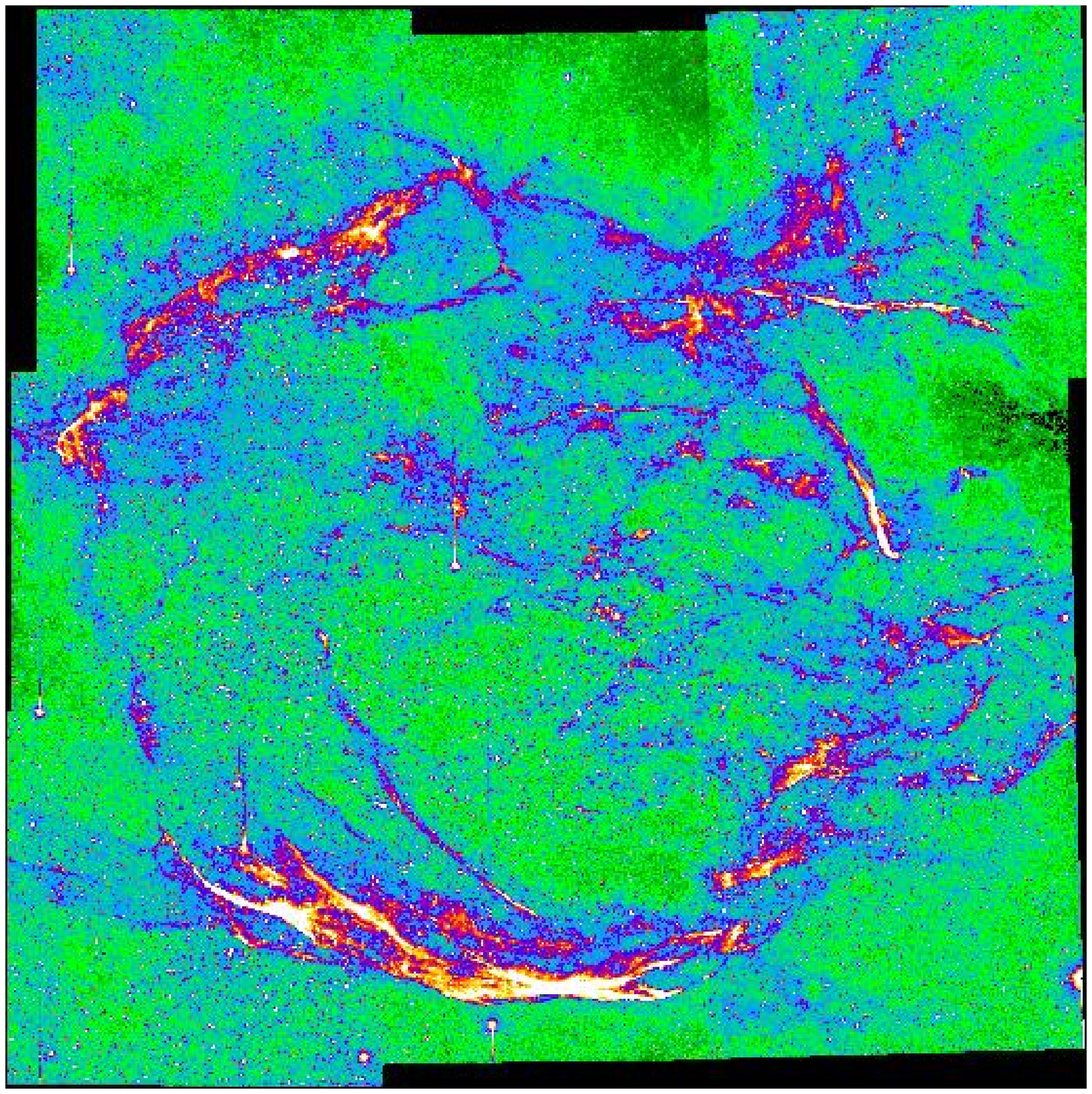}}
\\ {{\bf Figure 1.} The remnant \snr\ in the low ionization line of \oii 3727\AA\ }
\label{fig1}
\end{figure}

\section{Current preliminary results}
In Fig. 1 we show the \oii\ low ionization line image of \snr.
Several new structures are discovered in this field which were not known 
before. Eventhough the \oiii\ mosaic is under preparation the comparison of the
individual fields in \oii\ and \oiii\ reveals several significant morphological 
differences (Fig. 2). 
\par
The spectra analyzed up to now were taken in the south and south--west 
areas of the remnant and show the characteristic signature of emission from 
shock heated gas (\sii/\ha\ $>$ 1.0). 
In addition, these spectra display very strong
\oiii\ emission relative to \ha\ and, even stronger of course, relative 
to \hbeta. The measured values of the \oiii/\hbeta\ ratio lie in the 
range of $\sim$ 7--18, suggesting the presence of both complete and incomplete 
recombination zones. 
Furthermore, the prominent \oiii\ emission is indicative of shock speeds 
around $\sim$ 100 \vel\ (Fig. 2). We measure the  \ha/\hbeta\  
ratio around 5 showing that the interstellar extinction absorbs 
moderate amount of light from the remnant.
\begin{figure*}
\centering
\mbox{\epsfclipon\epsfxsize=2.8in\epsfbox[72 159 540 633]{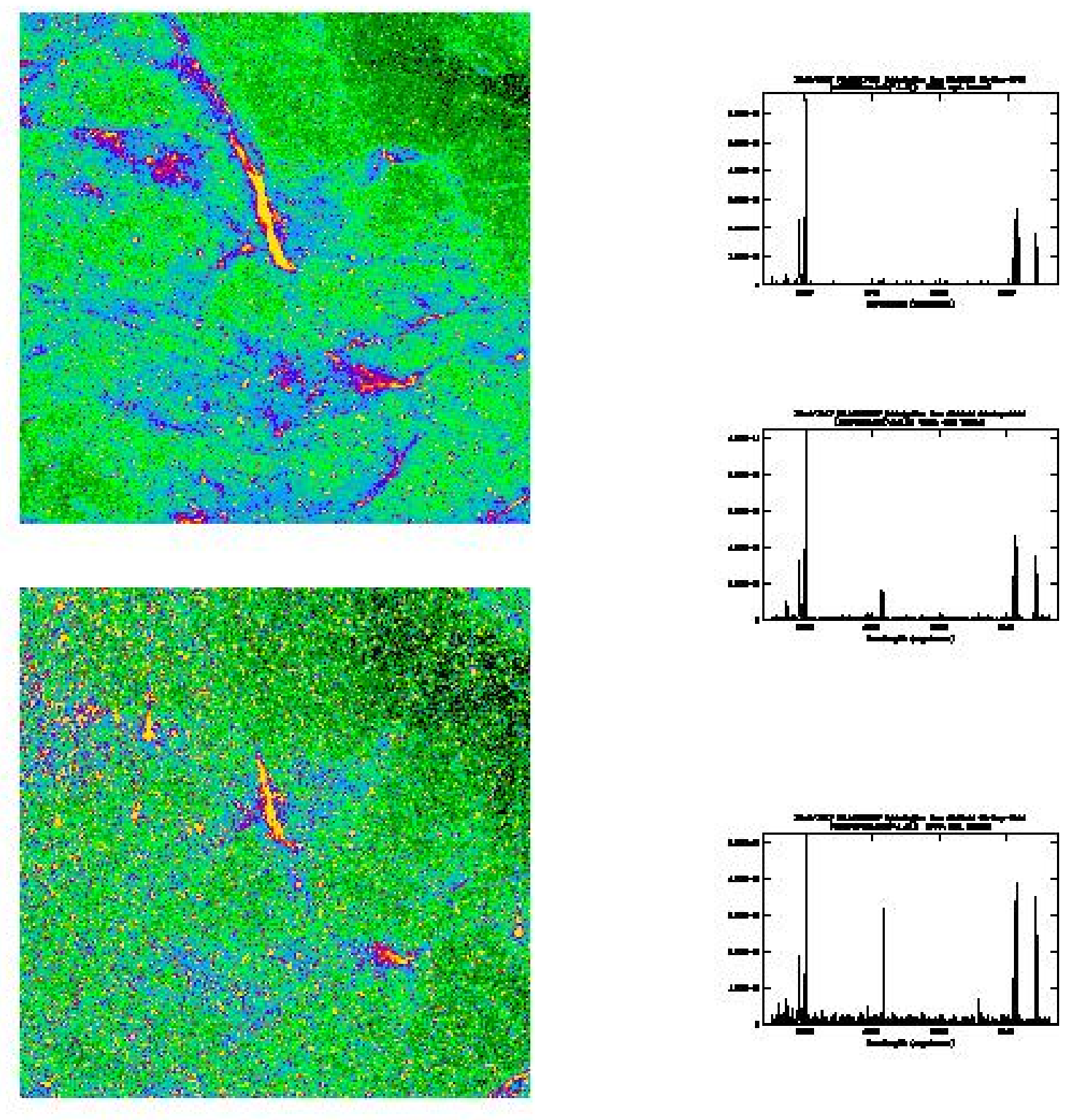}}
\\ {{\bf Figure 2.} A specific portion of the remnant in \oii\ 
(top left) and \oiii\ (bottom left) display different morphologies. The right
plot shows typical spectra of the remnant showing very strong \oiii\ emission.}
\label{fig2}
\end{figure*}
\begin{figure*}
\centering
\mbox{\epsfclipon\epsfxsize=2.8in\epsfbox[72 161 540 631]{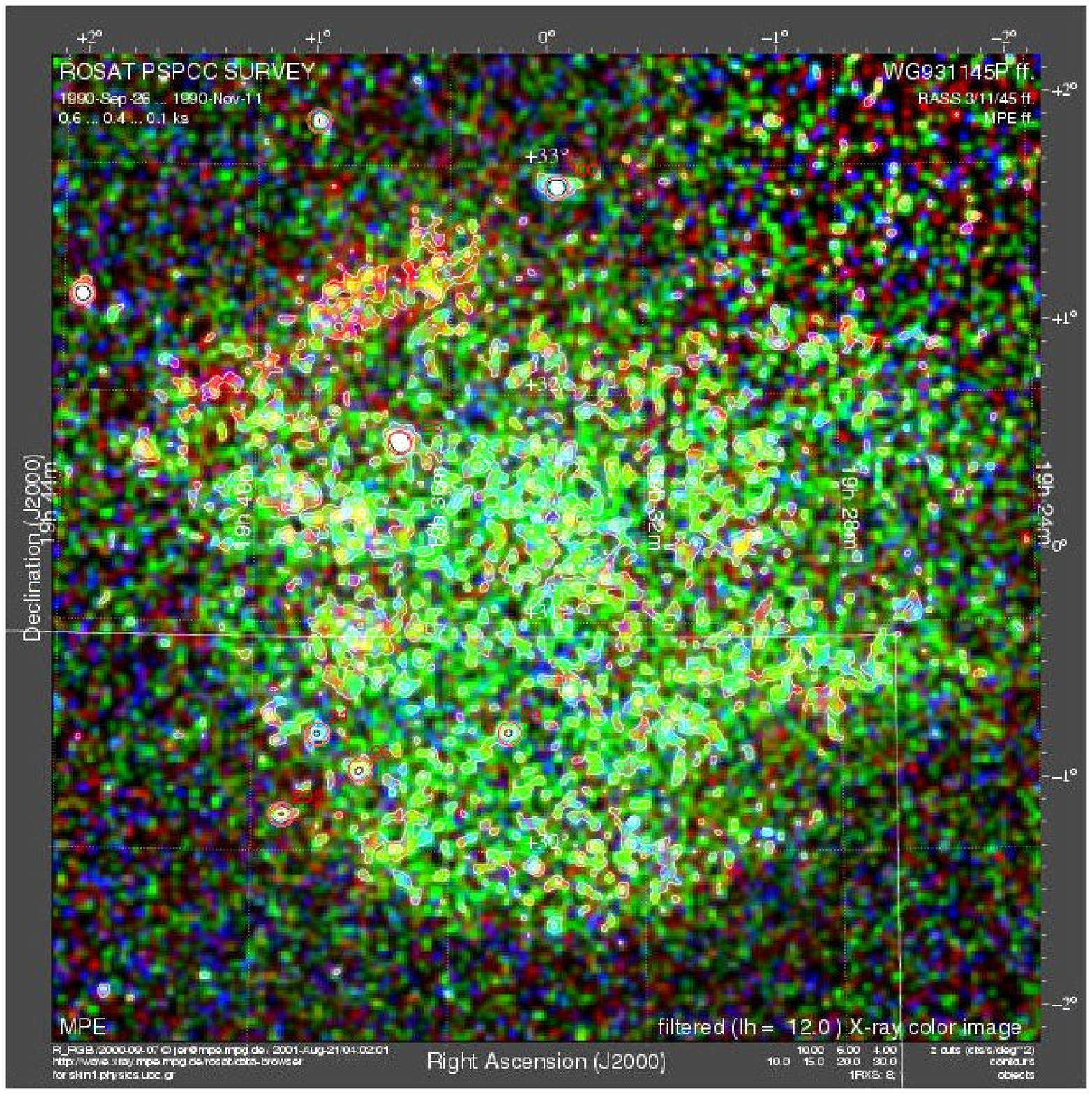}}
\\ {{\bf Figure 3.} The X--ray image of \snr\ in the 0.1--2.4 keV band 
observed by the ROSAT satellite during the All--Sky survey.}
\label{fig3}
\end{figure*}
\par
A quick search in the ROSAT All--Sky survey data base revealed weak soft X--ray 
emission from \snr\ (Fig. 3). It is evident that the X--ray emission is 
centrally peaked and is, roughly, bounded by the the outer \oii\ filaments. 
The survey photon event files were extracted and are under analysis in order 
to study in, as much detail as possible, the X--ray spectral properties of 
the remnant. 
\vfill\eject
\section*{Acknowledgments}
Skinakas Observatory is a collaborative project of the
University of Crete, the Foundation for Research and Technology-Hellas
and the Max-Planck-Institut fur Extraterrestrische Physik.

\section*{References}
Fesen R. A., Gull T. R., and Ketelsen D. A. 1983, ApJS 51, 337 \\
Fesen R. A., Blair W. P., and Kirshner R. P. 1985, ApJ 292, 29 \\
Gull T. R., Kirshner R. P., and Parker R. A. R. 1977, ApJ 215, L69 \\
Reich W., Berkhuijsen E. M., and Sofue Y. 1979, A\&A 72, 270 \\
Seward F. D. 1990, ApJS 73, 781\\

\end{document}